\documentclass[a4paper,12pt]{article}

\usepackage{amsmath}
\usepackage{amssymb}
\usepackage{latexsym}
\usepackage{cite}
\usepackage[dvips]{graphicx}
\usepackage{bbold}
\usepackage{epsfig}
\usepackage{amsfonts}

\newcommand{\beqa}{\begin{eqnarray}}
\newcommand{\eeqa}{\end{eqnarray}}
\newcommand{\bea}{\begin{eqnarray}}
\newcommand{\eea}{\end{eqnarray}}
\newcommand{\be}{\begin{equation}}
\newcommand{\ee}{\end{equation}}
\newcommand{\half}{\frac{1}{2}}
\newcommand{\tr}{\,\textup{tr}}

\newcommand{\mcA}{{\mathcal A}}

\newcommand{\mcF}{{\mathcal F}}

\newcommand{\mcM}{{\mathcal M}}

\newcommand{\mcO}{{\mathcal O}}

\newcommand{\mcK}{{\mathcal K}}
\renewcommand{\Re}{\textup{Re}}

\numberwithin{equation}{section}

\begin{document}
\title{\bf F-term Uplift in Heterotic M-theory\footnote{Partial results were presented at the 39th International Symposium Ahrenshoop.}}
\author{\normalsize{Filipe Paccetti Correia$^{\spadesuit}$\footnote{paccetti@fc.up.pt} \ and Michael G. Schmidt$^{\clubsuit}$\footnote{m.g.schmidt@thphys.uni-heidelberg.de}},\\
\\
{$^{\spadesuit}${\it \small{Centro de F\'\i sica do Porto}}},
\\ {{\it \small{Faculdade de Ci\^encias da Universidade do Porto}}},
\\ {{\it \small{Rua do Campo Alegre, 687, 4169-007 Porto, Portugal}}}\\
{$^{\clubsuit}${\it \small{Institut f\"ur Theoretische Physik}}},
\\ {{\it \small{Universit\"at Heidelberg}}},
\\ {{\it \small{Philosophenweg 16, 69120 Heidelberg, Germany}}}}

\date{\normalsize{May 6, 2009}}

\maketitle

\abstract{We investigate the viability of F-term uplift in heterotic M-theory. With this aim we explore a natural ingredient of heterotic compactifications, namely vector bundle moduli. It is shown that it is generically possible to obtain stable de Sitter vacua with broken supersymmetry provided the little K\"ahler potential and the prefactors of the non-perturbative superpotential are suitably tuned. An additional requirement is the existence of non-trivial gauge instantons both at the visible and hidden sectors. This is illustrated with analytical and numerical examples.}

\section{Introduction}

The search for stable string theory de Sitter vacua experienced a new boost in the last six years, triggered by the works of \cite{Giddings:2001yu,Kachru:2003aw}. Most of this effort has been directed to the type IIB string setting (for a recent review see \cite{Denef:2008wq}), and indeed led to new insights in string phenomenology and cosmology. It is highly desirable to extend these successes to the heterotic string/M theory too, for these naturally give rise to grand unified theories with a chiral spectrum. 

Swift progress towards moduli stabilization in the heterotic string/M-theory setting is hardened by a number of technical difficulties, of which we name a few: (i) absence of RR fluxes and non-K\"ahler\emph{ness} of the (6d) compactification manifold once the most general fluxes are taken into account; (ii) presence of non-trivial Hermitean Yang-Mills (HYM) instantons and the associated bundle moduli; (iii) lack of a truly compelling mechanism for uplifting anti de Sitter supersymmetric vacua to Minkowski or de Sitter ones. In practice, (i) means that after stabilising the complex structure by suitable NSNS fluxes, one still has a host of unfixed moduli. Then one faces two possibilities. The first is to rely on non-perturbative effects to stabilise the dilaton, the K\"ahler moduli and eventually also the vector bundle moduli. These non-perturbative effects are expected to be due to strong coupling dynamics (gaugino condensation) at the heterotic hidden sector and world-sheet/M2 brane instantons. This program has been pursued in the heterotic M-theory case in refs.\cite{Curio:2001qi,Curio:2003ur,Buchbinder:2003pi,Becker:2004gw,Buchbinder:2004im,Anguelova:2005jr,Braun:2006th,Curio:2006dc,Correia:2007sv}. A second possibility is to consider additional fluxes to stabilise the K\"ahler moduli \cite{deCarlos:2005kh,Anguelova:2006qf}, albeit at the price of rendering the 6d compactification manifold non-K\"ahler. Non-K\"ahler compactifications have the drawback that their low-energy descriptions are not as well understood as their Calabi-Yau counterparts. On the other hand, they do promise to be more flexible concerning the values that the flux superpotential can attain \cite{deCarlos:2005kh}. Combined with gaugino condensation this would also lead to stabilisation of the dilaton at phenomenologically acceptable values.

Some of the technical difficulties presented by HYM instantons (point (ii)) were addressed in our recent paper, ref.\cite{Correia:2007sv}, using a superfield formulation \cite{Paccetti:2004ri,Correia:2006pj} of 5d heterotic M-theory. There we explained how the bundle moduli enter the 4d K\"ahler potential and discussed how this influences moduli stabilisation in supersymmetric vacua. Continuing on this path, the aim of the present paper is to clarify in which circumstances the presence of bundle moduli can modify the vacuum energy and lead to supersymmetry breaking Minkowski or de Sitter vacua (point (iii)). Previously proposed uplift mechanisms for heterotic M-theory include uplift by charged matter VEVs \cite{Becker:2004gw}, D-term uplift \cite{Buchbinder:2004im}, S-track stabilisation due to the addition of putative M5 brane instantons \cite{Curio:2006dc}, and the inclusion of anti M5-branes in the 5d heterotic M-theory bulk \cite{Gray:2007qy}, but none of these has been so far completely worked out. In particular, none of these works addressed the impact of vector bundle moduli on the vacuum energy. It is the goal of the present work to contribute to close this gap.

\subsection{The underlying idea}\label{sec:idea}

The setup we consider in this paper is a \emph{minimal} truncation of heterotic M-theory \cite{Lukas:1998yy}, with $h^{1,1}=1$ K\"ahler modulus, and \emph{two} gauge singlet chiral superfields: one at the visible brane, another at the hidden one. The effective 4d K\"ahler and superpotential read \cite{Correia:2007sv}
\be\label{eq:K_pot_simple}
                  \mcK=\mcK(T+\bar{T}-k_0(\chi,\bar{\chi})-k_\pi(\phi,\bar{\phi});S_\pi+\bar{S}_\pi+\alpha k_\pi(\phi,\bar{\phi})) \ ,
\ee
where
\be\label{eq:K_pot_univ}
              \mcK(X,Y)=-3\ln\frac{3}{4\alpha}\left[(Y+\alpha X)^{\frac{4}{3}}-Y^{\frac{4}{3}}\right] \ ,
\ee
and
\be\label{eq:superpot}
                  W=W_{0}(\chi,\phi)+A(\chi,\phi)e^{-aT}+B(\phi)e^{-bS_\pi} \ .
\ee
Here, $T$ is the K\"ahler modulus, the so-called \emph{dilaton} $S_\pi$ is the Calabi-Yau volume modulus at the hidden brane, while $\chi$ and $\phi$ are vector bundle moduli, singlets under the unbroken gauge symmetries at the visible and hidden brane of heterotic M-theory, respectively. The K\"ahler potential in eqs.\eqref{eq:K_pot_simple} and \eqref{eq:K_pot_univ} follows from our superfield description \cite{Correia:2006pj,Correia:2007sv} of 5d heterotic M-theory. A more detailed explanation of these expressions will be provided in section \ref{sec:superpotential} and the appendix.

In the present paper, we want to explore the $(\chi,\phi)$-dependence of the \emph{little K\"ahler potentials}\footnote{The term "little K\"ahler potential" has been employed in the literature on type IIB flux compactifications to denote the K\"ahler potential for a D3 brane moduli space, which is the compactification manifold itself. In the present case, the little K\"ahler potential determines instead the geometry of the bundle moduli space.} $k_i$, and also $A(\chi,\phi)$, $B(\phi)$ and $W_0(\chi,\phi)$ in the superpotential above, to obtain supersymmetry breaking \emph{de Sitter} vacua. As we will discuss in section \ref{sec:superpotential}, while in general the non-perturbative prefactors are functions of the bundle moduli, for this to be also the case for $W_0$ one has to make an additional assumption, namely that by some mechanism certain charged scalars get non-vanishing VEVs. Avoiding this assumption implies that $W_0$ is a constant, and we might worry that it is then not possible to have enough uplift to obtain dS vacua. In fact, in type IIB KKLT like flux compactifications this seems to be the case (for explicit examples see \cite{Brown:2008zq}), and the reason for this is the no-scale structure of the K\"ahler potential.\footnote{In the case of multiple K\"ahler moduli the no-scale structure of the K\"ahler potential is not an obstacle to the existence of dS vacua \cite{Covi:2008ea}. However, the latter are still difficult to construct with well-motivated superpotentials \cite{Covi:2008zu}.} To be more precise, what we mean is that in the simplest type IIB setup with one K\"ahler modulus, gaugino condensation on a stack of D7 branes wrapping the 4-cycle, and a mobile D3 brane with position $z$, described by
\be\label{eq:typeIIB}
               \mcK=-3\ln(T+\bar{T}-k(z,\bar{z})) \ , \qquad W=W_0+A(z)e^{-aT} \ ,
\ee 
one cannot find stable dS minima \cite{GomezReino:2006dk,GomezReino:2006wv}. A way out of this problem, put forward in ref.\cite{Lebedev:2006qq} (see also \cite{GomezReino:2006dk}), is to allow the tree level superpotential $W_0$ to depend on additional hidden matter singlet(s). By contrast, since in the heterotic M-theory case the K\"ahler potential does not display a no-scale structure, we expect that it might be possible to have dS vacua even without invoking $W_{0}$ to be $\phi$-dependent. Likewise, as we will see in Sec. \ref{sec:toy}, replacing the K\"ahler potential of Eq.\eqref{eq:typeIIB} by $\mcK=-n\ln(T+\bar{T}-k(z,\bar{z}))$ with $n\geq 4$ is enough to allow for the construction of dS vacua by suitably choosing $k(z,\bar{z})$ and $A(z)$.

In the following we elaborate on the conditions needed for an uplift. In the case at hand, the scalar potential can be written as (see Appendix \ref{app:A})
\be
           V_F=V_0+\delta V_0 \ ,
\ee
where
\be
           V_0=e^\mcK\left[|W|^2+\mcK^{I\bar{J}}\partial_I\partial_{\bar{J}}|W|^2-(X^I+\bar{X}^I-k^I)(\partial_I+\partial_{\bar{I}})|W|^2\right] \ ,
\ee
and
\be\label{eq:F_uplift}
           \delta V_0=e^\mcK \mcK^{\chi\bar{\chi}}|k_\chi^I\partial_I W+\partial_\chi W|^2+e^\mcK \mcK^{\phi\bar{\phi}}|k_\phi^I\partial_I W+\partial_\phi W|^2 \ .
\ee
Here, $I$ runs only over the geometric moduli, $T$ and $S_\pi$ (i.e. $X^I=T,S_\pi$). We have split the potential into two terms. The first one, $V_0$, is not sensitive to the existence of the singlet $\phi$ in a significat way and leads to the AdS vacua studied in ref.\cite{Correia:2007sv}. The second term, $\delta V_0$, depends on the derivatives of the little K\"ahler potential and the superpotential with respect to $\phi$ and, therefore, can work as an uplift term. %A detailed computation (see Appendix \ref{app:A}) shows that in the present case
%\be
%          \mcK^{\phi\bar{\phi}}=\frac{k^{\phi\bar{\phi}}}{\alpha\mcK_{S_\pi}-\mcK_T} \ .
%\ee
%To determine the values of $\mcK_T$ and $\mcK_{S_\pi}$ we use the K\"ahler potential derived in \cite{Lalak:2001dv,Correia:2006pj}, which reads\footnote{This K\"ahler potential is obtained by a direct reduction of 5d heterotic M-theory \cite{Lukas:1998yy}. As such, it does not take into account possible corrections arising from higher order terms in 11d M-theory.}
%\be
%          \mcK=-3\ln\frac{3}{4\alpha}\left[(S_\pi+\bar{S}_\pi+\alpha(T+\bar{T}))^{\frac{4}{3}}-(S_\pi+\bar{S}_\pi)^{\frac{4}{3}}\right] \ .
%\ee
%Then, the uplifting term above can be rewritten as 
%\be\label{eq:F_uplift}
%          \delta V_0=\frac{1}{3}e^{\frac{2}{3}\mcK}\frac{k^{\phi\bar{\phi}}|k_\phi^I\partial_I W+\partial_\phi W|^2}{(S_\pi+\bar{S}_\pi)^{\frac{1}{3}}} \ .
%\ee
A necessary condition for the uplift to be non-zero is thus that at the minimum
\be\label{eq:uplift_cond}
          k_\alpha^I D_I W+D_\alpha W=k_\alpha^I\partial_I W+\partial_\alpha W\neq 0 \ ,
\ee
where $\alpha=\chi$ and/or $\phi$. Clearly, uplift thus necessarily implies supersymmetry breaking.

At first sight, the simplest way of obtaining enough uplift is to have $\partial_\phi W_0=\mu^2\neq 0$ and a neglegible contribution of non-perturbative terms to \eqref{eq:uplift_cond}. Then, by fine-tuning $\mu^2$ one can easily obtain enough uplift to render the vacuum energy positive. To avoid postulating a $\phi$-dependent tree-level superpotential, we might consider instead exploring the other  (exponentially suppressed) terms in \eqref{eq:uplift_cond}. Since in that case, the l.h.s. of this equation also depends on the geometric moduli $T$ and $S_\pi$, it is less clear that the uplift term can turn a negative energy vacuum into a positive energy one. What could go wrong is that the geometric moduli might conspire to suppress the l.h.s. of \eqref{eq:uplift_cond} in the \emph{new} vacuum. It is the purpose of this paper to investigate these matters in detail to find out when and how the uplift potential Eq.\eqref{eq:F_uplift} can lead to Minkowski and de Sitter vacua.

Section \ref{sec:superpotential} contains a discussion of all terms in the superpotential, with emphasis on the tree-level dependence on the bundle moduli, and of the K\"ahler potential, presented in Section \ref{sec:kaehler}, where we also comment on little K\"ahler transformations. Moduli stabilisation is the subject of the next two Sections. In Section \ref{sec:adS}, we investigate the feasibility of vacuum uplift without invoking a tree level bundle moduli dependence in the superpotential. In Section \ref{sec:mu_vacua} we explore such a term and give concrete numerical examples of stable Minkowski vacua. We close with a short discussion of future directions of research.

\vspace{12pt}

We conclude this introduction with two comments. The first one concerns the differences between our setup and the more phenomenological models discussed in \cite{Lowen:2008fm} and \cite{Jeong:2008ef} (based on \cite{Lebedev:2006qq}) in the context of heterotic string compactifications. The basic differences are in (i) the number of moduli involved in our problem; (ii) the form of the K\"ahler potential, in particular its dependence on the bundle modulus $\phi$. But most crucially, (iii) the fact that we do not need to invoke a tree-level dependence of the superpotential on $\phi$. In particular, unlike ref.\cite{Lowen:2008fm}, we do not assume the K\"ahler modulus $T$ to be stabilised at a \emph{self-dual} point and therefore we have to study a problem with at least three moduli. Also, the form of the K\"ahler potential in that reference is different from ours, leading to an uplifting term,
\be\nonumber
                \delta V_{LN}=e^{\mcK}|\partial_\phi W_0|^2 \ ,
\ee
which has no dependence on the K\"ahler modulus and a different dependence on the bundle modulus and dilaton, but is otherwise qualitatively similar.

We would like to comment also on the possibility of D-term uplifting \cite{Burgess:2003ic,Achucarro:2006zf}, which in the the context of heterotic M-theory has been discussed e.g. in \cite{Buchbinder:2004im,Correia:2007sv}. We consider an anomalous U(1)$_A$ acting at the hidden brane, the discussion for the visible brane being completely analogous. The axions of both the K\"ahler modulus $T$ and the hidden brane CY volume modulus $S_\pi$ are charged under this U(1)$_A$ symmetry. A computation of the induced D-term potential then leads to
\be
            V_D=e^{\frac{2}{3}\mcK}\frac{Q^2_A}{(S_\pi+\bar{S}_\pi)^{\frac{1}{3}}} \ ,
\ee 
where the charge $Q_A$ is determined by the first Chern class of the U(1)$_A$ gauge field. Since the form of this term is similar to the one of the F-term uplift discussed before (cf. Eq.\eqref{eq:F_uplift}), one might be led to consider also the possibility of D-term uplifting. However, as we pointed out in \cite{Correia:2007sv}, it is known that for a non-vanishing $Q_A$ (first Chern class of the U(1)$_A$) the prefactor $A(\phi)$ in the non-perturbative superpotential presented in \eqref{eq:superpot} vanishes. As such, it is not possible anymore to find stable vacua, AdS or else.

\section{The superpotential and K\"ahler potential}\label{sec:superpotential}

In this section we collect the K\"ahler potential and all superpotential terms that determine the F-term potential relevant for our discussion of stable Minkowski/de Sitter vacua. Special emphasis is placed in explaining the tree-level superpotential depence on bundle moduli.

\subsection{Tree level superpotential}

As we already emphasised, the \emph{low-energy} tree-level superpotential that is obtained in heterotic Calabi-Yau compactifications with hermitean Yang-Mills 
instantons does not contain bundle moduli self-couplings. In fact, this follows \cite{Witten:1985bz} from the holomorphicity condition $\mcF_{(0,2)}=0$ which is imposed on the background gauge fields by the requirement of supersymmetry (see e.g. discussion in \cite{Correia:2007sv}). On the other hand, the superpotential does depend on the bundle moduli through couplings to \emph{massive} charged modes, including Kaluza-Klein states. To generate an uplift term of the type discussed in \cite{Lowen:2008fm}, it is therefore crucial that some of these modes obtain non-vanishing Vev's. Let us explain this in more detail.

In heterotic M-theory, the tree-level superpotential can be obtained \cite{Correia:2007sv}\footnote{This is also true for non-K\"ahler heterotic compactifications, cf. \cite{Gurrieri:2007jg} for the half-flat case.} using the superpotential induced in the presence of a non-zero 4-form flux G,
\be
            W_{tree}=\int_{M_7} G\wedge\Omega \ ,
\ee
($M_7=CY_3\times S^1/Z_2$) where $\Omega$ is the Calabi-Yau's $(3,0)$-form. Then, a straightforward computation shows that the superpotential for the chiral superfields at the hidden brane is
\be
            W_{tree}\sim\int_{CY}\Omega\wedge\tr\left[\bar{\mcA}\wedge\bar{\partial}\bar{\mcA}+\frac{2}{3}\bar{\mcA}\wedge\bar{\mcA}\wedge\bar{\mcA}\right] \ ,
\ee
where $\bar{\mcA}=\bar{\mcA}_{\bar{n}}d\bar{z}^{\bar{n}}$ is the $(0,1)$-component of the gauge field at the hidden brane. The 4d superpotential couplings are obtained by expressing the gauge field as a sum of a background $\bar{\mcA}(\phi)$ configuration and charged fluctuations, which we denote collectivelly by $C$:
\be
            \bar{\mcA}=\bar{\mcA}(\phi)+C \ .
\ee 
The background configuration $\bar{\mcA}(\phi)$ is parametrized by a number of \emph{gauge bundle} moduli, singlets under the unbroken gauge symmetries, and must be holomorphic in the sense that the $(0,2)$-component of the gauge field strength must vanish, $\mcF_{(0,2)}(\phi)=0$. Then, one can show that, up to a constant piece, the superpotential reads
\be\label{eq:sup_pot_C}
            W_{tree}\sim\int_{CY}\Omega\wedge\tr\left[C\wedge\bar{D}C+\frac{2}{3}C\wedge C\wedge C\right] \ ,
\ee
where 
\be
            \bar{D}C\equiv \bar{\partial}C+C\wedge\bar{\mcA}+\bar{\mcA}\wedge C \ .
\ee
The cubic term on the r.h.s. of \eqref{eq:sup_pot_C} leads to the low-energy Yukawa couplings. In particular, in the standard embedding case we obtain in this way the standard ${\bf 27}$ Yukawa couplings. Massless charged modes satisfy $\bar{D}C=0$ in which case the first term in \eqref{eq:sup_pot_C} vanishes. Of interest to us in this paper, therefore, are \emph{non-zero} modes for which
\be
            \bar{D}C\neq 0 \ ,
\ee
as in this case, the first term in the superpotential can lead to mass terms\footnote{These massive fields include both Kaluza-Klein states as well as states with masses below the Kaluza-Klein scale.} \cite{Kuriyama:2008pv} for charged matter of the type
\be
                      W\sim m(\phi) C^2 \ .
\ee

It is a well known feature that, as one moves in the vector bundle moduli space parameterized by the $\phi^\alpha$, the number of massless charged states $n_C$ can change. Generically, there are hypersurfaces in moduli space where this number increases. Then, once one departs from the hypersurface a number of these states becomes massive. This phenomenon is described by the above superpotential as follows. There must be a sub-matrix of $m_{ab}(\phi)$ that vanishes at the hypersurface which, by a suitable field redefinition, we set to be at $\phi=0$.\footnote{For simplicity, in this paper we are assuming the hypersurface to be of (complex) codimension one.} As $\phi\neq 0$, that sub-matrix of the KK mass matrix becomes non-zero, and the corresponding charged states become massive. It is possible to describe in more detail what is happening near these hypersurfaces. In the following, we will denote the massive states we are interested on as $Q$:
\be
            \bar{\mcA}=\bar{\mcA}(\phi)+\bar{u}_Q Q + \cdots \ .
\ee
These are defined by $(0,1)$-forms $\bar{u}_Q$ satisfying the eigenvalue equation
\be\label{eq:laplace}
                     \bar{D}^{\dagger}\bar{D}\,\bar{u}_Q=M^2(\phi,\bar{\phi})\bar{u}_Q \ ,
\ee 
where $\bar{D}^{\dagger}\equiv-*D*$ is the adjoint of $\bar{D}$. At the locus $\phi=0$ where the $Q$ become massless we have $M^2=0$, and $\bar{u}_Q=\bar{u}_Q^0$ is an element of Doulbeault cohomology (at $\phi=0$),
\be
                     \bar{D}_0\bar{u}_Q^0=0 \ .
\ee
Expanding to first order around $\phi=0$, 
\be
                     \bar{u}_Q=\bar{u}_Q^0+\phi\,\bar{u}_Q^1+\mcO (\phi^2) \ ,
\ee
we find
\be
                     \phi\,\bar{D}_0^{\dagger}\bar{D}_0\,\bar{u}_Q^1=\left(\phi\partial_\phi M^2(0)+\bar{\phi}\partial_{\bar{\phi}} M^2(0)\right)\bar{u}_Q^0 \ .
\ee
Then, the fact that $M^2$ is a real function of $\phi$ and $\bar{\phi}$ implies that
\be
              \bar{D}_0^{\dagger}\bar{D}_0\,\bar{u}_Q^1=0 \ ,
\ee
and therefore, to first order in the bundle modulus $\phi$, $\bar{u}_Q$ is an element of the $\bar{D}_0$-Dolbeault cohomology.
\be
             \bar{D}_0\bar{u}_Q=0 \ .
\ee
Armed with this knowledge, we can compute the superpotential mass term near the locus $\phi=0$. In fact, it follows from this equation that
\be
             \bar{D}\bar{u}_Q=\phi\left(\partial_\phi\bar{\mcA}(0)\wedge\bar{u}_Q+\bar{u}_Q\wedge\partial_\phi\bar{\mcA}(0) \right) + \mcO(\phi^2) \ ,
\ee
and plugging this back in the superpotential \eqref{eq:sup_pot_C} we find the following term
\be
             W=\phi \,m_{ab}Q^aQ^b \ ,
\ee
with
\be
                    m_{ab}\sim \int_{CY}\Omega\wedge\tr\left(\partial_\phi\bar{\mcA}(0)\wedge\bar{u}_{Q^a}\wedge\bar{u}_{Q^b}\right) \ ,
\ee
a function of the complex structure and bundle moduli. Concrete examples of this type of couplings have been constructed e.g. in \cite{Bouchard:2006dn,Braun:2006em,Braun:2006da} with an eye on realizing dynamical SUSY breaking or generating $\mu$-terms in heterotic M-theory. Then, eventually, the \emph{meson} $Q^T{\bf m}Q$ dynamically acquires a non-vanishing Vev,
\be
                            \langle Q^T{\bf m}Q\rangle\neq 0 \ .
\ee
To realize the proposal of \cite{Lebedev:2006qq} in the heterotic M-theory setting, later in sec.\ref{sec:mu_vacua} we will have to assume this to be the case. The detailed mechanism leading to this Vev will not be of concern to us. Relevant for us will be the fact that the classical superpotential contains a term linear in $\phi$
\be\label{eq:tree_sup}
                        W_{tree}=\mu^2\,\phi+\cdots 
\ee
and therefore $\partial_\phi W_{tree}\neq 0$ at the locus $\phi=0$.

\subsection{Non-pertubative superpotential}

Non-perturbative contributions to the superpotential of heterotic M-theory have been discussed in several places in the literature. They can arise due to M2-brane instantons \cite{Curio:2001qi,Moore:2000fs,Lima:2001jc} stretching between the boundary branes and wrapping holomorphic cycles, and due to gaugino condensation \cite{Dine:1985rz,Kaplunovsky:1993rd,Brignole:1993dj} taking place at the hidden brane \cite{Horava:1996vs,Nilles:1997cm,Lalak:1997zu,Lukas:1997rb,Lukas:1999kt}. In our case, with $h^{1,1}=1$, there is only one cycle that the M2-brane instanton can wrap, and the non-perturbative part of the superpotential reads
\be
                   W_{np}=A(\phi,\,z)\,e^{-a\,T}+B(\phi,\,z)\,e^{-b\,S_\pi} \ ,
\ee 
where $A(\phi,\,z)$ and $B(\phi,\,z)$ are one-loop determinants for the classical M2-brane instanton and the gauge instanton contributions, respectively. Generically, these prefactors depend both on the bundle moduli $\phi$ and the complex structure moduli $z$. Specific examples of this dependence for M2 brane instantons have been computed \cite{Buchbinder:2002ic,Buchbinder:2002pr,Curio:2008cm} for vector bundles over elliptically fibered Calabi-Yau 3-folds described by the \emph{spectral cover} construction. In these cases, the prefactor $A(\phi,\, z)$ is a homogeneous polynomial function of the bundle moduli. As far as we know, the dependence of the prefactor $B(\phi,\,z)$ on the bundle moduli has not been discussed in the literature so far. 

%For simplicity, in this paper we take it to be proportional to a suitable power of $A(\phi,\, z)$:
%\be
%                    B(\phi,\,z)=C(z) \left(A(\phi,\, z)\right)^{-\frac{b\alpha}{a}} \ ,
%\ee
%where $C(z)$ depends on the complex structure moduli. In general, $C$ could also depend on bundle moduli $\chi$ not associated with deformations of the vector bundle restricted to holomorphic curve wrapped by the M2-brane instanton. However, for the sake of calculability, in this paper we assume the existence of only one bundle modulus.

\subsection{The K\"ahler potential}\label{sec:kaehler}

In \cite{Correia:2007sv}, we showed that the dependence of the K\"ahler potential on the bundle moduli is determined by $h^{1,1}$ so-called little K\"ahler potentials
\be\label{eq:little_k_I}
k^I_i(\phi_i,\bar{\phi}_i)=i\lambda\int_{CY}\tr(\mcA_i(\bar{\phi}_i)\wedge\bar{\mcA}_i(\phi_i))\wedge\omega^I \ ,
\ee
as
\be
            \mcK=\mcK(T^I+\bar{T}^I-k^I_0-k^I_\pi;S_\pi+\bar{S}_\pi+\alpha_Ik^I_\pi) \ .
\ee 
Here, the $(2,2)$-forms $\omega^I$ span the CY's $H^{2,2}$ cohomology and the $\alpha^I$ are the instanton numbers. The $\phi_i$ stand for all bundle moduli, both at the visible brane ($i=0$) and the hidden brane $i=\pi$. In this paper we consider the simplest case leading to uplift, i.e. one bundle modulus at each brane. While this small number of bundle moduli is unrealistic, our results can straightforwardly be extended to the many moduli case. Also, for the sake of computational simplicity, we take $h^{1,1}=1$. Then $\omega^I\to J\wedge J$ in Eq.\eqref{eq:little_k_I}, where $J$ is the CY's K\"ahler 2-form. In the definition \eqref{eq:little_k_I} of the little K\"ahler potentials we are implicitly assuming that there is a \emph{gauge} for which the $(0,1)$-component of the gauge connection, $\bar{\mcA}$, depends only on the holomorphic bundle moduli $\phi_i$. 

For $h^{1,1}=1$, in \cite{Correia:2006pj,Correia:2007sv} we were able to determine the \emph{exact} K\"ahler potential that follows from a reduction to 4d of the 5d heterotic M-theory of ref.\cite{Lukas:1998yy}. It reads
\be
            \mcK=-3\ln\frac{3}{4\alpha}\left[(S_\pi+\bar{S}_\pi+\alpha(T+\bar{T})-\alpha k_0)^{\frac{4}{3}}-(S_\pi+\bar{S}_\pi+\alpha k_\pi)^{\frac{4}{3}}\right] \ .
\ee
In the weak coupling limit $\Re(S_\pi)\gg\alpha\Re(T)$, we then find
\be\label{eq:kaehler_weak}
            \mcK\simeq -\ln(S_\pi+\bar{S}_\pi+\alpha k_\pi)-3\ln(T+\bar{T}-k_0-k_\pi) \ ,
\ee
as expected. Later, in sections \ref{sec:adS} and \ref{sec:mu_vacua} we will study vacua with $\alpha\Re(T)\lesssim \Re(S_\pi)/3$, for which \eqref{eq:kaehler_weak} still is a very good approximation.

%Near the $\phi=0$ locus described in the previous section, the K\"ahler potential can be brought to the form of Eq.\eqref{eq:K_pot_simple}, i.e. the little K\"ahler potential is
%\be\label{eq:lit_k_simple}
%                   k(\phi,\bar{\phi})\simeq|\phi|^2 \ .
%\ee
%Note that in writing this we are neglecting all charged matter and additional vector bundle moduli but the general case is a simple generalization. As above, we write the gauge connection as
%\be
%                  \bar{\mcA}(\phi)\simeq \bar{\mcA}(0)+ \bar{u}_\phi\,\phi \ ,
%\ee
%where $\bar{u}_\phi=\partial_\phi\bar{\mcA}(0)$. Plugging this in \eqref{eq:little_k} we find
%\be
%                 k(\phi,\bar{\phi})\simeq k_0+(f_\phi\, \phi+\textup{h.c.})+k_{\phi\bar{\phi}}|\phi|^2 \ ,
%\ee
%with
%\be
%                 f_\phi=i\lambda\int_{CY}\tr(\mcA(0)\wedge\bar{u}_\phi)\wedge J\wedge J \ ,
%\ee
%and
%\be
%                 k_{\phi\bar{\phi}}=i\lambda\int_{CY}\tr(u_{\bar{\phi}}\wedge\bar{u}_\phi)\wedge J\wedge J \ .
%\ee

The full K\"ahler potential is invariant under so-called \emph{little} K\"ahler transformations,
\be
                T\to T+g(\phi) \ ,
\ee
\be
                S_\pi\to  S_\pi -\alpha g(\phi) \ ,
\ee
\be
                k(\phi,\bar{\phi})\to k(\phi,\bar{\phi})+g(\phi)+ \bar{g}(\bar{\phi})\ .
\ee
%Thus, we see that by means of a little K\"ahler transformation with
%\be
%                g(\phi)=\half k_0 + f_\phi \phi{app:A} \ ,
%\ee
%one can define away the constants $k_0$ and $f_\phi$ to obtain a quadratic little K\"ahler potential. 
Note that under little K\"ahler transformations also the $\phi$-dependent prefactors of the non-pertubative terms in the superpotential \eqref{eq:superpot} get transformed. Indeed, the transformation is as follows:
\be
                A(\phi)\to A(\phi)e^{-ag(\phi)} \ ,
\ee
\be
                B(\phi)\to B(\phi)e^{b\alpha g(\phi)} \ .
\ee
These transformations have a (microscopical) geometric origin and are related to the fact that both the axions and the little K\"ahler potential(s) are defined up to gauge transformations. Then, the 1-loop prefactors $A(\phi)$ and $B(\phi)$ should be seen as sections of line bundles, $\mcO_A(\mcM)$ and $\mcO_B(\mcM)$, over the vector bundle moduli space $\mcM$. The curvatures of $\mcO_A(\mcM)$ and $\mcO_B(\mcM)$ should both be proportional to $\partial\bar{\partial}k(\phi,\bar{\phi})$. As such, it could be that powers of both sections are related to each other as $B(\phi)\propto \left(A(\phi)\right)^{-\frac{b\alpha}{a}}$, but we do not assume this to be the case.

\section{Stable de Sitter vacua without matter VEVs}\label{sec:adS}

In a previous paper \cite{Correia:2007sv} we showed how in heterotic M-theory with the superpotential Eq.\eqref{eq:superpot} and constant $W_0$ and $B$, supersymmetric AdS vacua can always be found. These features are present if the gauge bundle at the hidden brane is trivial and the matter sector does not acquire non-trivial VEVs. In this case, the supersymmetry condition determining the value of $\chi$ simplifies considerably, reading
\be\label{eq:susy_visible}
                \partial_\chi A(\chi)-ak_{0,\chi} A(\chi)=0 \ .
\ee
Once we embed a non-trivial bundle at the hidden sector, $B$ becomes a function of the associated bundle moduli $\phi$. In this case, in addition to to replacing $A(\chi)$ by $A(\chi,\phi)$ in \eqref{eq:susy_visible}, supersymmetry also leads to
\be\label{eq:susy_hidden}
                (\partial_\phi A(\chi,\phi)-ak_{\pi,\phi} A(\chi,\phi))e^{-aT}+(\partial_\phi B(\phi)+\alpha\, b\, k_{\pi,\phi} B(\phi))e^{-bS_\pi}=0 \ .
\ee
In the case mentioned at the end of the previous section, that $B(\phi)$ is proportional to a certain power of $A(\phi)$, we find that the supersymmetry condition Eq.\eqref{eq:susy_hidden} reduces to Eq.\eqref{eq:susy_visible}.

As explained in sec.\ref{sec:idea}, once supersymmetry is broken, certain terms in the potential become non-vanishing and positive and thus might act as uplift terms. It is therefore natural to wonder if it is possible to setup the prefactors $A(\chi,\phi)$ and $B(\phi)$, as well as the little K\"ahler potentials $k_0(\chi,\bar{\chi})$ and $k_\pi(\phi,\bar{\phi})$, in order to obtain de Sitter vacua (with broken supersymmetry). The answer to this question is positive, but to a certain degree obtaining this result relies on numerical trial and error. To make our findings more transparent for the reader, we will illustrate the main features of this mechanism with the help of a \emph{toy model} before focusing the heterotic M-theory case.

\subsection{A toy model}\label{sec:toy}
Our toy model is defined by
\be
                  \mcK=-n\ln(T+\bar{T}-k(z,\bar{z})) \ ,
\ee
\be
                  W=W_0+A(z)e^{-aT} \ .
\ee
For $n=3$ this model describes the situation one encounters in type IIB KKLT like compactifications, with a no-scale structure ($\mcK^{i\bar{j}}\mcK_i\mcK_{\bar{j}}=3$). For $n=4$ we find instead a situation in several regards similar to the one we discuss in this paper, for which $\mcK^{i\bar{j}}\mcK_i\mcK_{\bar{j}}=4$, too. 

Our analysis is simplified by redefining $T\to T+\frac{1}{a}\ln(A(z)/A_0)$ and $k\to \tilde{k}+\ln|A(z)/A_0|^2$. Then,
\be
                  \mcK=-n\ln(T+\bar{T}-\tilde{k}(z,\bar{z})) \ , \qquad W=W_0+A_0e^{-aT} \ .
\ee
Taking $\partial_z \tilde{k}=0$ necessarily implies that no Minkowski stable vacuum can be achieved. To see this, we note that in the limit that the model can be trusted, i.e. for $a\Re(T)\gg 1$, the potential can be arranged as
\be
                  V_0\simeq\frac{|W_0|^2}{(T+\bar{T}-\tilde{k})^n}\left[-3+\frac{1}{n}\left|\Phi+n\right|^2\right] \ ,
\ee 
where
\be\label{eq:defphi}
                  \Phi\equiv a(T+\bar{T}-\tilde{k})(A_0/W_0)e^{-aT} \ .
\ee
To derive this result we used the following set of relations (see Appendix \ref{app:A}),
\be
                   \mcK^{T\bar{T}}=\frac{1}{n}(T+\bar{T}-\tilde{k})^2-\frac{k^{z\bar{z}}|\tilde{k}_z|^2}{\mcK_T} \ ,
\ee
\be
                   \mcK^{T\bar{z}}=-\frac{k^{z\bar{z}}\tilde{k}_z}{\mcK_T} \ , \qquad \mcK^{z\bar{z}}=-\frac{k^{z\bar{z}}}{\mcK_T} \ .
\ee

A Minkowski vacuum ($V=\partial_{\bar{\Phi}}V=0$) would then imply both that $\Phi\simeq-n$ and $V=0$, which is of course not possible. This situation might change however if, at the vacuum, $\partial_z k\neq 0$. We can always set $z=0$ at the vacuum and, by a suitable redefinition of $T$, take $\tilde{k}(z=0)=0$ and
\be
                  \tilde{k}(z,\bar{z})=c\,z+\bar{c}\bar{z}+|z|^2+\cdots
\ee 
Then, at $z=0$ the potential becomes
\be\label{eq:pot_toy}
                 V\simeq V_0+\frac{a^2\,|c|^2}{n}\frac{|A_0|^2e^{-a(T+\bar{T})}}{(T+\bar{T})^{n-1}}\propto -3+\frac{1}{n}\left|\Phi+n\right|^2+\frac{|c|^2}{n\,(T+\bar{T})}|\Phi|^2 \ .
\ee
Imposing the Minkowski vacuum conditions, we then find, again to leading order in an expansion in inverse powers of $a\Re(T)$, that 
\be\label{eq:Phi_1}
                \Phi\simeq -\frac{n}{1+\frac{|c|^2}{T+\bar{T}}} \ ,
\ee
and (with $V=0$)
\be\label{eq:Re(T)}
                \Re(T)\simeq \frac{n-3}{6}|c|^2=\frac{n-3}{6}k^{z\bar{z}}|\tilde{k}_z|^2 \ ,
\ee
for $n\neq 3$. Plugging this back in Eq.\eqref{eq:Phi_1}, we obtain
\be
               \Phi\simeq -(n-3) \ 
\ee
which fixes the value of the modulus $T$ given $a,\,n$ and $(A_0/W_0)$. The amount of uplift, i.e. the value of $|c|^2$, needed to obtain a Minkowski vacuum is then given by \eqref{eq:Re(T)}. For $n=3$, we find instead
\be
                \Re(T)\simeq -\half\left(|c|^2+\frac{6}{a}\right)<0 \ . 
\ee
Moreover, since 
\be
                \partial_\Phi\partial_{\bar{\Phi}}V=\frac{|W_0|^2}{n(T+\bar{T})^n}\left(1+\frac{|c|^2}{T+\bar{T}}\right)>0 \ ,
\ee
and $\partial_\Phi^2 V$ is subleading, we conclude that for $n>3$ Minkowski vacua are possible while for $n\leq 3$ the opposite is the case. 

One last condition still has to be imposed on the derivatives of $\tilde{k}(z,\bar{z})$ at $z=0$, to ensure that at this point the potential really has a minimum with respect to $z$ too. This can be achieved, for example, if one takes $c=\bar{c}$ and
\be\label{eq:ansatz}
                  \tilde{k}(z,\bar{z})=c\,z+c\bar{z}+|z|^2-\frac{n}{n-3}(z^2+\bar{z}^2) \ ,
\ee 
leading to
\be
                  V\simeq \frac{3|W_0|^2}{n(T+\bar{T})^{n+1}}\left(3(n-3)|z|^2-z^2-\bar{z}^2\right) \ ,
\ee
which for $n\geq 4$ has a \emph{minimum} at $z=0$. The $z$-modulus thus gets stabilised with a mass of order
\be
                   m^2_z\sim \frac{|W_0|^2}{(T+\bar{T})^{n}} \ .
\ee 

One can also check that with the Ansatz \eqref{eq:ansatz} there is a supersymmetric AdS vacuum, for which $z=c(n-3)/(n+3)$ is unequal to zero, and the value of $T$ in the vacuum is given by solving
\be
                   \Phi=-\frac{n}{1+\frac{n}{a(T+\bar{T})}} \ ,
\ee
but we will not further dwell on this issue.

Closing this section, we would like to comment on the effect that considering a larger number of $z$-moduli could have on the uplift mechanism. This is relevant in the context of heterotic string/M-theory compactifications, where one typically has dozens of bundle moduli. This possibility can be taken into account by modifying the parameter $|c|^2$ that determines the amount of uplift as
\be
                   |c|^2=k^{\bar{j}i}k_ik_{\bar{j}} \ .
\ee
Unless there is some special (unknown to us) structure constraining this quantity in the bundle moduli space, as there is for the K\"ahler moduli space, it is reasonable to believe that increasing the number of moduli also means increasing the amount of uplift due to the combined effect of all bundle moduli. That would mean that the typically large number of bundle moduli should make it easier to obtain dS vacua in heterotic M-theory.

\subsection{Results for heterotic M-theory model}\label{sec:main}

We now come back to the main object of investigation in this paper, i.e. the heterotic M-theory model introduced in section \ref{sec:idea}, with a constant tree-level superpotential ($\mu^2=0$). Recall that the main differences with respect to the toy model discussed above lie in the number of moduli (now two geometric moduli and two bundle moduli - one per brane) and in the more complicated form of the uplift terms appearing in the uplift potential \eqref{eq:F_uplift}. This last aspect is a relevant one, as we now explain using a reasoning in the line of the previous section. We introduce the variables
\be\label{eq:defphi2}
                  \Phi\equiv a(T+\bar{T}-k_0-k_\pi)(A/W_0)e^{-aT} \ ,
\ee
and
\be\label{eq:defpsi}
                  \Psi\equiv b(S_\pi+\bar{S}_\pi+\alpha k_\pi)(B/W_0)e^{-bS_\pi} \ .
\ee
Using suitable little K\"ahler transformations one can always set the $k_i$ to zero in the vacuum, and we will do this here. Then, for $a\Re(T)\gg 1$ and $b\Re(S_\pi)\gg 1$, the potential before the uplift can be approximated as (cf. Appendix \ref{app:A})
\be
                  V_0\simeq e^\mcK |W_0|^2\left[-3+\frac{1}{3}|\Phi+3|^2+|\Psi+1|^2\right] \ .
\ee
Following the same type of argument as above, we find that at the minimum $\Phi\simeq -3$ and $\Psi\simeq -1$, leading to \emph{adS} vacua. To this potential we now add the uplift terms, which, again neglecting a number of subleading contributions, we write as (cf. Eq.\eqref{app:deltaV} and following)
\be
                 V\simeq V_0+e^\mcK |W_0|^2\left[\frac{|c|^2}{3(T+\bar{T})}|\Phi|^2+\frac{T+\bar{T}}{3}\left|d\frac{\Phi}{T+\bar{T}}+e\frac{\Psi}{S_\pi+\bar{S}_\pi}\right|^2\right] \ ,
\ee
where
\be
                 |c|^2=k^{\chi\bar{\chi}}|\partial_\chi k_0-\tfrac{1}{a}\partial_\chi\ln A|^2 \ ,
\ee
\be
                 d=\sqrt{k^{\phi\bar{\phi}}}\partial_\phi(k_\pi-\tfrac{1}{a}\ln A) \ ,
\ee
\be
                 e=-\sqrt{k^{\phi\bar{\phi}}}\partial_\phi(\alpha k_\pi+\tfrac{1}{b}\ln B) \ .
\ee

An interesting observation is that, if we take $e=0$ we retrieve the same potential for $\Phi$ as in the toy model with $n=3$ (cf. eq.\eqref{eq:pot_toy}), which we showed to lead to insufficient uplift. On the other hand a straightforward but lengthy calculation shows also that setting $c=0$ is not compatible with a Minkowski or dS vacuum. Thus, we find that it is necessary to have at least one bundle modulus at each brane and $c,e\neq 0$. 

We focus now on the particular case with $d=0$. For Minkowski vacua we find
\be
          \Phi=-\frac{3}{1+\frac{|c|^2}{T+\bar{T}}} \ ,\quad \Psi=\frac{1}{3}\left(\frac{|c|^2}{T+\bar{T}}-2\right)\Phi \ ,
\ee
and
\be\label{eq:from_e_to_c}
              9=\left(\frac{|c|^2}{T+\bar{T}}-2\right)\frac{|e|^2(T+\bar{T})}{(S_\pi+\bar{S}_\pi)^2} \ .
\ee
At these vacua (with $d=0$), the mass matrix of the geometric moduli can also be easily computed. We obtain
\be
              V_{T\bar T}\simeq 3a^2\,\frac{m_{3/2}^2}{1+\frac{|c|^2}{T+\bar{T}}} \ ,
\ee
and
\be
              V_{S_\pi\bar{S}_\pi}\simeq   \frac{b^2}{3a^2}\left(\frac{|c|^2}{T+\bar{T}}-2\right) V_{T\bar T} \ ,
\ee
while all other components of the matrix ($V_{TT}$, $V_{T\bar{S}_\pi}$, etc.) are subleading. Clearly, for $|c|^2>4\Re(T)$ these vacua are minima with respect to the geometric moduli. The precise values of their masses depends on the form of the little K\"ahler potentials and superpotential prefactors, a knowledge that we do not have. We are however able to determine \emph{lower bounds} on their values. We find
\be
              m_T^2\geq \mcK^{T\bar T}V_{T\bar T}\simeq m_{3/2}^2\frac{a^2(T+\bar T)^2}{1+\frac{|c|^2}{T+\bar{T}}} \ ,
\ee
\be
              m_{S_\pi}\geq \mcK^{S_\pi\bar{S}_\pi}V_{S_\pi \bar{S}_\pi}\simeq\frac{b^2(S_\pi+\bar{S}_\pi)^ 2}{a^2(T+\bar T)^2}\left(\frac{|c|^2}{T+\bar{T}}-2\right) \mcK^{T\bar T}V_{T\bar T} \ .
\ee

It is also relevant to investigate the supersymmetry breaking F-terms,
\be
                  F^I=e^{\mcK/2}|W|\mcK^{I\bar{J}}{\bar{D}_{\bar J}\bar{W}}/{\bar{W}} \ .
\ee
As in the case of the mass spectrum, we are not able to determine the values of the F-terms without a precise knowledge of the little K\"ahler potentials and superpotential prefactors. However, assuming that $k_\chi\sim(\partial_\chi\ln A)/a$ and $\alpha k_\phi\sim(\partial_\phi \ln B)/b$, we can estimate
\be\label{eq:Fterms1}
              F^{S_\pi}\simeq f^{S_\pi}+\mcO_1\, m_{3/2} \ ,
\ee
\be\label{eq:Fterms2}
              F^T\simeq f^{T}+\mcO_2\, m_{3/2} \ ,
\ee
where
\be\label{eq:Fterms3}
              f^{S_\pi}=3\frac{S_\pi+\bar{S}_\pi}{|c|^2}f^T \ , \quad f^T=-m_{3/2}\frac{|c|^2}{1+\frac{|c|^2}{T+\bar{T}}} \ ,
\ee
while the unknown terms are of order
\be
              \mcO_1\sim |e|^2\bar{\Phi} \ , \, |e|^2\bar{\Psi}\frac{T+\bar T}{S_\pi+\bar{S}_\pi} \ ,
\ee
and
\be
              \mcO_2\sim |c|^2\bar{\Phi} \ , \, |e|^2 \bar{\Psi}\frac{T+\bar T}{S_\pi+\bar{S}_\pi} \ , \, |e|^2 \ .
\ee

\begin{table}
\begin{center} 
\begin{tabular}[t]{|c|c|c|c|}
\hline&&\\[-2ex]
$|A_0|$ & $10^{-5}$ & $1$ & $10^{-5}$
\\[1ex]
$|B_0|$ & $2.6\cdot10^{-7}$ & $0.7\cdot 10^{-7}$ & $1.3\cdot 10^{-7}$
\\[1ex]
$b$ & $5$ & $5$ & $30$
\\[1ex]
$|W_0|$ & $2.6\cdot 10^{-6}$ & $0.7\cdot 10^{-6}$ & $0.8\cdot 10^{-14}$
\\[1ex]\hline\hline &&\\[-2ex]
$c$ & $0.36$ & $0.85$ & $1.1$
\\[1ex]
$e$ & $3.6$ & $1.7$ & $4.1$
\\[1ex]
$\Re(T)$ & $0.03$ & $0.18$ & $0.25$
\\[1ex]
$\Re(S_\pi)$ & $0.97$ & $0.82$ & $0.75$
\\[1ex]
$m_{3/2}$ & $1.2\cdot 10^{-4}$ & $2.7\cdot 10^{-6}$ & $1.8 \cdot 10^{-14}$
\\[1ex]
$m_{S}$ & $\gtrsim 0.88~m_{3/2}$ & $\gtrsim 0.95~m_{3/2}$ & $\gtrsim 3.7~m_{3/2}$
\\[1ex]
$m_{T}$ & $\gtrsim 3.7~m_{3/2}$ & $\gtrsim 20~m_{3/2}$ & $\gtrsim 27~m_{3/2}$
\\[1ex]
$f^{T}$ & $-4.3\cdot 10^{-2}m_{3/2}$ & $-0.24~m_{3/2}$ & $-0.35~m_{3/2}$
\\[1ex]
$f^{S}$ & $45~m_{3/2}$ & $6.8~m_{3/2}$ & $3.7~m_{3/2}$
\\[1ex]\hline
\end{tabular}
\end{center} 
\caption{A sample of Minkowski vacua with $\mu^2=0$, $a=100$, $\alpha=1$, $d=0$. Masses are measured in units of $M_P$.
\label{tb:num_results_0}}
\end{table}

Aiming to discuss the physical implications of these expressions we will first set some notation. We introduce $A_0,B_0$ as the values of $A(\chi,\phi)$ and $B(\phi)$ at the vacuum. Following \cite{Buchbinder:2003pi} we suitably rescale the moduli in such a way that at the phenomenologically interesting vacuum $\Re(S_0)\equiv \Re(S_\pi)+\alpha\Re(T)=1$.\footnote{The modulus $\Re(S_0)$ measures the size of the CY 3-fold at the visible brane.} Then, we have $a\sim 10^2$ and, depending on the hidden brane unbroken gauge group, $b\sim 5-30$. The value of $B$ is also gauge group (and bundle moduli) dependent, but should be of order $|B_0|\sim 10^{-6}-10^{-7} M_P^3$. Finally, the instanton number should be of order $\alpha\sim 1$. 

In Table \ref{tb:num_results_0} we present a few physical features of a sample of three different Minkowski vacua with $d=0$. Note that we allowed ourselves to tune the constant part of the superpotential $W_0$ at quite different values, depending on the value of $b$. For $b\sim 5$ we have a value of $W_0$ of more or less\footnote{The value of $W_0$ is determined by order of magnitude arguments. The authors of \cite{Gukov:2003cy} have pointed out that this value can be lowered by a few orders of magnitude by considering fractional Chern-Simons invariants on CY 3-folds with non-trivial fundamental group.} the natural order of magnitude \cite{Buchbinder:2003pi}. By contrast, for the value $b\sim 30$ that leads to a TeV scale gravitino mass, $W_0$ is by far too low and we have no explanation for its value. The value of $\Re(T)$ in the third example presented in Table \ref{tb:num_results_0} is near an upper bound set by the different scales involved in the problem. This feature is shared with the vacua with $\mu^2\neq 0$ that we will find in the following section, to which we postpone a more detailed discussion of this bound.

In all three examples presented in Table \ref{tb:num_results_0} we find $|f^{S_\pi}|\gg |f^T|$. However this does not necessarily imply that supersymmetry breaking is dilaton dominated ( i.e. that $|F^{S_\pi}|\gg |F^T|$). In fact, in the third example we have $|e|^2\gg 1$, which means that the unknown terms $\mcO_i$ introduced in Eqs. \eqref{eq:Fterms1} and \eqref{eq:Fterms2} might well dominate over the $f^I$ presented in Table \ref{tb:num_results_0}. In that case $F^T$ and $F^{S_\pi}$ could be of the same order. By contrast, in the first example presented in the table, we do have $f^{S_\pi}\gg \mcO_i$ so that we can state that supersymmetry breaking is dilaton dominated. Note that given $A_0,B_0,W_0,a,b$, fixing $\Re(S_0)=1$ determines both $|c|$ and $|e|$. This means that for $d=0$ the just mentioned features of supersymmetry breaking are fixed. It is however not excluded that the additional freedom introduced by considering non-vanishing values of $d$ can modify this situation.

Finally let us mention that we do not provide values for the masses of the bundle moduli $\phi$ and $\chi$ as these are determined by unknown coefficients in the little K\"ahler potentials. But we should emphasize that, as it was the case for the toy model, it is always possible to choose these coefficients as to ensure that we are at minima of the potential with respect to the bundle moduli too.

\section{Stable de Sitter vacua with matter VEVs}\label{sec:mu_vacua}

We have just shown that it is not necessary to invoke a tree-level dependence of the superpotential on bundle moduli to find Minkowski or dS vacua in heterotic M-theory. For the sake of completeness, however, we will also investigate this possibility here. Our framework is different from that in \cite{Lowen:2008fm,Jeong:2008ef}, the differences lying in the number of geometric moduli (which is smaller in ref.\cite{Lowen:2008fm}) and detailed form of the K\"ahler potential and superpotential, but our findings are qualitatively similar.

We thus consider a tree-level term in the superpotential of the form (cf. \eqref{eq:tree_sup})
\be
                 W_{tree}=W_0+\mu^2\phi \ .
\ee
We further assume that unlike it was the case in the previous section, $c,~d$ and $e$ are neglegible at the vacuum. Then, in contrast to that case, for $a(T+\bar{T})\gg 1$ the positions of the geometric moduli in the vacuum are not significantly altered by a value of $\mu^2\lesssim |W_0|$, only the the vacuum energy is affected. It is then not difficult to find out that at the vacuum
\be
                 V_0\simeq -3|W_0|^2e^\mcK \ ,
\ee
and thus, to obtain a Minkowski vacuum on has to set
\be  
                    |\mu|^4\simeq 3\mcK_{\phi\bar{\phi}}|W_0|^2 \ ,
\ee
for we have $\delta V_0\simeq e^\mcK \mcK^{\phi\bar{\phi}}|\mu|^4$. Since the uplift disturbs the position of the minimum only very weakly, the supersymmetry conditions can be used to gain some qualitative insight on the features of Minkowski and dS vacua. From $\Phi\simeq -3$ and $\Psi\simeq -1$ we find
\be\label{eq:qualitat}
              |W_0|\sim a\Re(T)|A e^{-aT}|\sim b\Re(S_\pi)|Be^{-bS_\pi}| \ . 
\ee
We can use these '\emph{order of}' relations to set upper bounds to the values of both $\Re(T)$ and the Planck scale $M_{P}$. Crucially, these bounds cannot be removed by the same effect that uplifts the vacuum. The following discussion applies also to the cases discussed in the previous section.

\begin{table}
\begin{center} 
\begin{tabular}[t]{|c|c|c|c|}
\hline&&\\[-2ex]
$|A|$ & $10^{-6}$ & $1$ & $10^{-5}$
\\[1ex]
$|B|$ & $10^{-6}$ & $10^{-6}$ & $2.5\cdot 10^{-7}$
\\[1ex]
$b$ & $5$ & $5$ & $30$
\\[1ex]
$|W_0|$ & $1.57\cdot 10^{-7}$ & $2.9\cdot 10^{-7}$ & $2.7\cdot 10^{-15}$
\\[1ex]\hline\hline &&\\[-2ex]
$|\mu|^2$ & $7.146\cdot 10^{-7}$ & $7.026\cdot 10^{-7}$ & $5.535\cdot 10^{-15}$
\\[1ex]
$\Re(T)$ & $0.032$ & $0.177$ & $0.25$
\\[1ex]
$\Re(S_\pi)$ & $0.965$ & $0.812$ & $0.751$
\\[1ex]
$m_{3/2}$ & $4.8\cdot 10^{-6}$ & $8.9\cdot 10^{-7}$ & $4.96 \cdot 10^{-15}$
\\[1ex]
$m_{S}$ & $5.4~m_{3/2}$ & $5.4~m_{3/2}$ & $53~m_{3/2}$
\\[1ex]
$m_{T}$ & $7.2~m_{3/2}$ & $41~m_{3/2}$ & $46~m_{3/2}$
\\[1ex]
$m_{A}$ & $(0.2;~0.14)m_{3/2}$ & $(11;~1.8)m_{3/2}$ & $(25;~23)m_{3/2}$
\\[1ex]
$F_{T}$ & $-1.3\cdot 10^{-2}m_{3/2}$ & $1.6\cdot 10^{-2}m_{3/2}$ & $-1.8\cdot 10^{-2}m_{3/2}$
\\[1ex]
$F_{S}$ & $-0.5~m_{3/2}$ & $-0.7~m_{3/2}$ & $0.12~m_{3/2}$
\\[1ex]\hline
\end{tabular}
\end{center} 
\caption{Numerical examples with $\mu^2\neq 0$, $a=100$, $\alpha=1$. Masses are measured in units of $M_P$.
\label{tb:num_results}}
\end{table}

\begin{figure}
\begin{center}
\resizebox{0.6\textwidth}{!}{
  \includegraphics{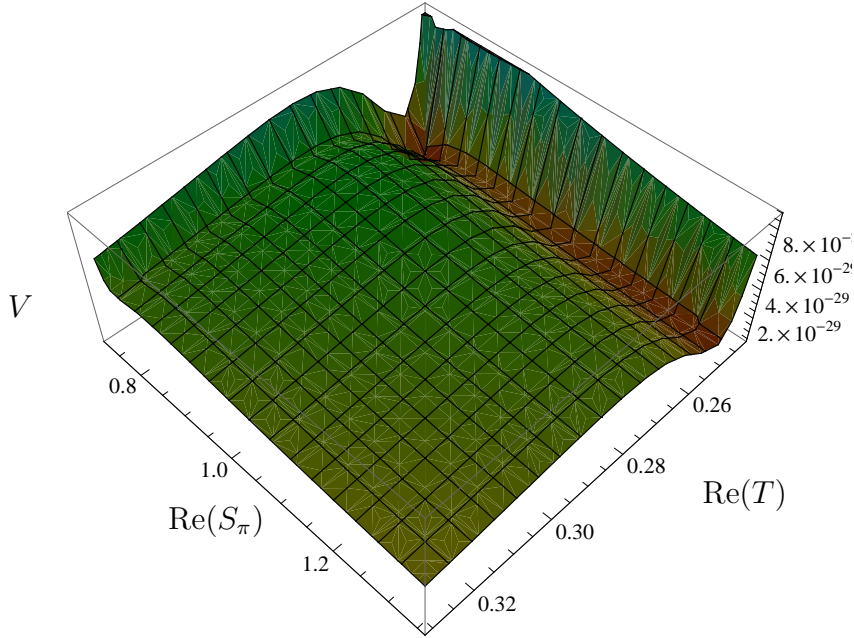}
}
%\vspace{5cm}       % Give the correct figure height in cm
\put(-260,40){$\Re(S_\pi)$}
\put(-60,50){$\Re(T)$}
\put(-320,120){$V$}
\caption{Uplifted F-term potential as a function of $\Re(T)$ and $\Re(S_\pi)$.}
\label{fig:1}       % Give a unique label
\end{center}
\end{figure}

Assuming as before that we are in the phenomenologically interesting vacuum  with $\Re(S_0)=1$, one can investigate if it is possible to \emph{tune} $W_0$ and $A$ to obtain the correct value for $M_P$. Using the above relations we find 
\be
              m_{3/2}\sim e^{\mcK/2}\frac{|W_0|}{M_P^2}\sim \frac{|A|}{M_P^2}e^{-a\Re(T)} \ .
\ee
We thus see that $\Re(T)$ is bounded by the ratio of the gravitino mass  to the Planck mass:
\be
              \Re(T)\lesssim \frac{1}{a}\ln\left(\frac{M_P}{m_{3/2}}\right) \ ,
\ee
where we used that $|A|\lesssim M_P^3$. Since $a\sim 10^2$ or larger, it is clear that
\be
              \Re(T)\lesssim 0.4 \ .
\ee 
From this, we then find that fixing the GUT scale at its phenomenological value, we always obtain a value for the Planck mass which is at least a bit too low
\be
              \langle M_P^2 \rangle\lesssim 0.4 M_P^2 \ ,
\ee
where $M_P^2$ is the measured Plank scale and $\langle M_P^2 \rangle$ is the value we obtain in our vacuum. The upper bound is attained for a gravitino mass of order the weak scale, $m_{3/2}\sim 10^{-16}M_P$, and $|A|\sim M_P^3$. The former can be achieved for $b\sim 30$, which is naturally obtained when the gauge instanton at the hidden brane breaks the $E_8$ down to a smaller gauge group. On the other hand, the value of $A$ depends on the bundle modulus $\phi$, but it is usually not expected that it takes values of order $M_P^3$. In fact, decreasing the value of $|A|$ has the effect of further decreasing the value of $\langle M_P^2 \rangle$. A more conservative estimate of $|A|\sim 10^{-5}$ (see \cite{Becker:2004gw}) leads instead to the upper bound
\be
              \Re(T)\lesssim 0.25 \to \langle M_P^2 \rangle\lesssim 0.25 M_P^2 \ .
\ee

These findings are confirmed by a numerical search for stable Minkowski vacua with $\mu^2\neq 0$. In table \ref{tb:num_results} we present the results of this search for \emph{three} distinct cases. Note that, for simplicity, by a suitable little K\"ahler transformation we set $k_{\phi\bar\phi}=1$ and $k_\phi=0$. In all cases, supersymmetry breaking is \emph{dilaton} dominated, i.e. $|F_{S_\pi}|\gg |F_T|$. We also present a graphical illustration of the uplifted potential as a function of $\Re(S_\pi)$ and $\Re(T)$ in Fig.\ref{fig:1}. To produce this figure, we set all other fields to their local minima.

\section{Conclusions and Outlook}

In this paper we achieved stabilisation in Minkowski vacua of a minimal version of heterotic M-theory with one K\"ahler modulus and up to two bundle moduli, one per brane, in addition to the dilaton. Since both uplift mechanisms discussed herein rely on the existence of vector bundle moduli, we had to assume a non-standard embedding of the gauge instanton. This, of course, nicely fits the fact that to obtain suitable heterotic GUTs one typically has to go beyond the simplest possibility of setting the gauge connection equal the spin connection. As we pointed out, it is not necessary to invoke the \emph{perturbative} tree-level part of the superpotential to depend on the bundle modulus to obtain sufficient uplift: The non-perturbative terms might well do the job provided one has non-trivial gauge instantons embedded in both heterotic M-theory branes. This is a particularity of heterotic M-theory that would not work in no-scale models.

Due to the complexity involved in constructing heterotic compactifications, it is far from clear if it is possible to find concrete realizations of both mechanisms discussed in this paper. There has been some progress in finding closed expressions for both the tree-level \cite{Anderson:2009ge} and the non-perturbative M2 brane instanton superpotentials (see \cite{Curio:2008cm} and references therein). Unfortunately, we are far from being able to determine the K\"ahler potential for bundle moduli explicitly for any non-trivial non-abelian HYM instanton over a smooth CY three-fold. It is therefore unclear if the values of order one that we took for $|c|^2$ and $|e|^2$ in section \ref{sec:main} are possible to attain. However, as we already pointed out, even if their values turn out to be one or two orders of magnitude smaller, this could be again compensated by the cumulative effect of a large number of bundle moduli, which is typically found to be in the order of the dozens.

Possible extensions of the framework discussed in this paper migth help avoiding the tension between a phenomenologically acceptable value of the GUT scale and the correct value of the Planck scale. In particular, the introduction of additional K\"ahler moduli and M5 branes in the bulk is a possible line of continuation of this work, which we are presently investigating.

Having discussed moduli stabilisation and uplift within a supergravity approximation it is also a natural next step to investigate the possibility of achieving slow-roll inflation in the same framework. We will consider this possibility in a future publication.

\section*{Acknowledgements}
We thank A. Lukas and J. Gray for useful discussions. F.P.C. is greatful both to the Institut f\"ur Theoretische Physik, Heidelberg, Germany, and the Rudolf Peierls Centre for Theoretical Physics, Oxford, U.K., for support and hospitality during visits in several stages of this work. The work of F.P.C. is supported by \emph{Funda\c c\~ao para a Ci\^encia e a Tecnologia} through the grant SFRH/BPD/20667/2004.

\appendix

\section{Useful formulae}\label{app:A}

The effective K\"ahler potential for heterotic M-theory has a number of interesting properties \cite{Correia:2007sv} which we now describe. Up to a term depending on the complex structure moduli, the K\"ahler potential is of the form
\be
            \mcK=\mcK\left(T^I+\bar{T}^I-k_0(\chi,\bar{\chi};C_0,\bar{C}_0)-k_\pi^I(\phi,\bar{\phi};C_\pi,\bar{C}_\pi);S_\pi+\bar{S}_\pi+\alpha_Ik_\pi^I(\phi,\bar{\phi};C_\pi,\bar{C}_\pi)\right) \ .
\ee 
Here, the geometric moduli are $h^{1,1}$ K\"ahler moduli $T^I$ and the hidden brane CY$_3$ volume modulus $S_\pi$. Descending from the 10d boundary Yang-Mills gauge fields we have a number of vector bundle moduli $\chi^\alpha$ and charged matter $C^a_0$ at the visible brane, vector bundle moduli $\phi^\alpha$ and charged matter $C^a_\pi$ at the hidden brane. These enter the K\"ahler potential through the little K\"ahler potentials 
\be\label{eq:little_k}
           k^I_i=i\lambda\int_{CY}\tr(\mcA_i\wedge\bar{\mcA}_i)\wedge\omega^I \ ,
\ee
with $\bar{\mcA}_0=\bar{\mcA}(\chi,C_0)$ (similar for $i=\pi$), and $\{\omega^I\}$ a basis of the $H^{2,2}$ cohomology. Finally, the instanton numbers $\alpha_I$ are determined by the second Chern classes of the Calabi-Yau (tangent bundle) and the gauge bundles at the boundary branes.

The fact that $\exp(-\mcK/3)$ is an homogeneous function of degree four of the geometric moduli for $k^I=0$, implies that 
\be
               \mcK_I(X^I+\bar{X}^I-k^I)=-4 \ ,
\ee
\be
               \mcK_{I\bar{j}}(X^I+\bar{X}^I-k^I)=-\mcK_{\bar{j}} \ ,
\ee
and therefore
\be
               \mcK^{i\bar{j}}\mcK_{\bar{j}}=-\delta^i_I(X^I+\bar{X}^I-k^I) \ ,
\ee
and
\be
                      \mcK^{i\bar{j}}\mcK_i\mcK_{\bar{j}}=4 \ .
\ee
This can then be used to rewrite the F-term potential (after the complex structure has been stabilised) as
\be
              V=e^\mcK\left[|W|^2+\mcK^{i\bar{j}}\partial_i\partial_{\bar{j}}|W|^2-(X^I+\bar{X}^I-k^I)(\partial_I+\partial_{\bar{I}})|W|^2\right] \ .
\ee
To further evaluate this expression we express the inverse K\"ahler metric $\mcK^{i\bar{j}}$ in terms of the inverse of $\mcK_{IJ}$, which we call $\mcK_0^{IJ}$, and the inverse of $\mcK_{0,\alpha\bar{\beta}}\equiv-\mcK_Ik^I_{\alpha\bar{\beta}}$, denoted by $\mcK_0^{\alpha\bar{\beta}}$:
\be
              \mcK^{IJ}=\mcK_0^{IJ}+\mcK_0^{\alpha\bar{\beta}}k_\alpha^I k_{\bar{\beta}}^J \ ,
\ee
\be
              \mcK^{I\bar{\beta}}=\mcK_0^{\alpha\bar{\beta}}k_\alpha^I \ ,
\ee
\be
              \mcK^{\alpha\bar{\beta}}=\mcK_0^{\alpha\bar{\beta}} \ .
\ee
For the single geometric modulus case this result reduces to the one obtained in \cite{Burgess:2006cb}, in the context of type IIB compactifications with D3-branes. 

In the minimal truncation of heterotic M-theory considered in this paper, the low energy spectrum contains two geometric moduli: one K\"ahler modulus $T$ and the CY$_3$ (hidden brane) volume modulus $S_\pi$. Also, for simplicity, we consider the situation that there is only one vector bundle modulus per brane. Then, the F-term potential can be written as
\be
           V_F=V_0+\delta V_0 \ ,
\ee
where
\be
           V_0=e^\mcK\left[|W|^2+\mcK^{I\bar{J}}\partial_I\partial_{\bar{J}}|W|^2-(X^I+\bar{X}^I-k^I)(\partial_I+\partial_{\bar{I}})|W|^2\right] \ ,
\ee
and
\be\label{app:deltaV}
           \delta V_0=e^\mcK \mcK^{\chi\bar{\chi}}|k_\chi^I\partial_I W+\partial_\chi W|^2+e^\mcK \mcK^{\phi\bar{\phi}}|k_\phi^I\partial_I W+\partial_\phi W|^2 \ .
\ee
Here, $I$ runs only over the geometric moduli, $T$ and $S_\pi$ (i.e. $X^I=T,S_\pi$), and
\be
           k^T=k_0(\chi,\bar{\chi})+k_\pi(\phi,\bar{\phi}) \ , \qquad k^{S_\pi}=-\alpha k_\pi(\phi,\bar{\phi}) \ ,
\ee
while
\be
           \mcK^{\chi\bar{\chi}}=-\mcK_T^{-1}k_0^{\chi\bar{\chi}} \ , \qquad \mcK^{\phi\bar{\phi}}=-(\mcK_T-\alpha\mcK_{S_\pi})^{-1}k_\pi^{\phi\bar{\phi}} \ .
\ee
Finally, we specify the K\"ahler potential
\be
              \mcK=-3\ln\frac{3}{4\alpha}\left[(S_\pi+\bar{S}_\pi+\alpha(T+\bar{T})-\alpha k_0(\chi,\bar{\chi}))^{\frac{4}{3}}-(S_\pi+\bar{S}_\pi+\alpha k_\pi(\phi,\bar{\phi}))^{\frac{4}{3}}\right] \ ,
\ee
to obtain 
\be
              -\mcK_T=3e^{\mcK/3}(S_\pi+\bar{S}_\pi+\alpha(T+\bar{T})-\alpha k_0)^{\frac{1}{3}} \ ,
\ee
and
\be
              -\mcK_T+\alpha \mcK_{S_\pi}=3e^{\mcK/3}(S_\pi+\bar{S}_\pi+\alpha k_\pi)^{\frac{1}{3}} \ .
\ee

\end{document}